\documentclass[prb,aps,floats,twocolumn,showpacs,groupedaddress,floatfix,amsmath]{revtex4}

\def\bc{\begin{center}}
\def\ec{\end{center}}
\def\bw{\begin{widetext}}
\def\ew{\end{widetext}}

\renewcommand{\vec}[1]{\mbox{\boldmath$#1$}}
\usepackage{amssymb}
\usepackage{graphicx}
\usepackage{dcolumn}
\usepackage{bm}
\usepackage{subfigure}
\usepackage{array}
\usepackage{multirow}
\usepackage{booktabs}
\usepackage{appendix}
\usepackage{nicefrac}

\begin{document}

\title{Microscopic Study of Edge Excitations of Spin-Polarized and Spin-Unpolarized $\nu=2/3$ Fractional Quantum Hall Effect}
\author{Ying-Hai Wu, G. J. Sreejith, and Jainendra K. Jain}
\affiliation{Department of Physics, The Pennsylvania State University, University Park, PA 16802}

\date{\today}

\begin{abstract}
The edge of spin unpolarized or spin polarized $\nu=2/3$ fractional quantum Hall states is predicted by the effective theory to support a backward moving neutral mode in addition to a forward moving charge mode. We study this issue from a microscopic perspective where these states are identified with effective filling factor of 2 of composite fermions, but with an effective magnetic field that is antiparallel to the external field. A simple counting from the composite fermion description suggests that there might be {\em two} backward moving edge modes, but explicit calculations show that one of these is projected out of the low energy sector, while the remaining mode provides a good microscopic account of the actual counter propagating edge mode. The forward moving modes are identified as ``Schur modes," obtained by multiplying the ground state wave function by the symmetric Schur polynomials. The edge of the 2/3 spin unpolarized state provides a particularly striking realization of ``spin charge separation" in a one dimensional Tomonaga Luttinger liquids, with the spin and charge modes moving in opposite directions.   
\end{abstract}

\maketitle

\section{Introduction}

Two-dimensional electron systems have been the platform for many interesting phenomena. In particular, the integer [\onlinecite{IQHE}] and fractional [\onlinecite{FQHE}] quantum Hall effects occur when a two-dimensional electron system (2DES) is placed in a magnetic field. Integer quantum Hall (IQH) states occur when an integer number of Landau levels are completely filled with electrons. For a partially filled Landau level (LL), interactions between electrons can produce incompressible states at certain fillings and lead to fractional quantum Hall (FQH) states. These are characterized by the formation of composite fermions [\onlinecite{JainCF}], where a composite fermion (CF) is the bound state of an electron and an even number of vortices. A strongly interacting state of electrons in a magnetic field $B$ is described by a weakly interacting state of composite fermions in an effective magnetic field $B^*$, whose direction can be either parallel or antiparallel to $B$. The composite fermions form Landau-like levels (called $\Lambda$ levels) in the field $B^*$, in analogy to the LLs of non-interacting electrons. The FQH states of electrons are described as IQH states of composite fermions, which correspond to situations where composite fermions occupy an integer number of $\Lambda$ levels. This results in FQH effect at the prominently observed fractions 
\begin{equation}
\nu = \frac{n}{2pn \pm 1}
\label{fillfactor}
\end{equation} 
where $+ (-) $ indicate that the direction of the effective magnetic field is parallel (antiparallel) to the real magnetic field. 

Since FQH states occur in the presence of a large magnetic field, one might at first expect that the spin degree of freedom is frozen. However, in the most widely studied GaAs system, the $g$-factor is very small, and unpolarized or partially polarized quantum Hall states have been found to occur. The CF theory predicts the possible spin polarizations at various fractions in terms of composite fermions filling both up and down spin $\Lambda$ levels ($\Lambda$Ls) [\onlinecite{Wu93,Park98}]. The spin polarization is determined by a competition between the CF cyclotron energy and the Zeeman energy $E_Z=g{\mu_B}B$; at very small Zeeman energies the state with smallest spin polarization is obtained, and transitions into larger spin polarizations occur as the Zeeman energy is increased. This physics has been found to be in good qualitative and semi-quantitative agreement with experiments [\onlinecite{SpinDegreeExp}]. In particular, both the 2/5 and the 2/3 FQH states map into filling factor 2 of composite fermions (with effective magnetic field antiparallel to the applied field for 2/3); the spin unpolarized state maps into the state in which 0$\uparrow$ and 0$\downarrow$ $\Lambda$Ls are occupied, and the fully polarized state is described as the one in which 0$\uparrow$ and 1$\uparrow$ $\Lambda$Ls are occupied.

Our concern in this paper is with the physics of the edge excitations of the FQH states. The FQH states are gapped in the bulk but there are gapless excitations residing at the boundary [\onlinecite{FQHEdge}], which provide a realization of a nontrivial one dimensional Tomonaga-Luttinger liquid [\onlinecite{WenReview,ChangReview}]. Several theoretical approaches have been used to study the edge states, especially the Chern-Simons theory [\onlinecite{WenReview,Levitov}]. In general, the FQH state at $n/(2pn+1)$ has $n$ edge modes, one corresponding to each $\Lambda$L. A surprising prediction of the edge theory has been the presence of backward moving neutral modes for the FQH states at $n/(2pn-1)$ for which the effective magnetic field for composite fermions is antiparallel to the real magnetic field [\onlinecite{WenReview,FradkinSpin,Kane}].  Evidence of such counter-propagating edge modes has been seen in a recent experiment [\onlinecite{TwoThirdNeutral}].

We will consider the FQH state at 2/3, which is the simplest state where backward moving modes are theoretically predicted. The bulk physics of both the spin unpolarized and spin polarized FQH states at 2/3 is closely related to that of the spin unpolarized and spin polarized IQH states at 2, as has been demonstrated by Wu, Dev and Jain [\onlinecite{Wu93}]. In this paper, we study the edge states of $\nu=2/3$ using a combination of the parton method, exact diagonalization, and the microscopic CF theory to test predictions of the effective field theory, and also to gain further insight into the physics of the backward moving edge modes. 

The presence of up-stream modes can be motivated in different ways. 
For fully spin polarized states, the presence of such modes appears naturally for the FQH states at 
\begin{equation}
\nu=1-\frac{n}{2pn+1},
\end{equation}
which are particle hole symmetric to the principal states. Consider the fully spin polarized 2/3 FQH state for example, which can be viewed as the 1/3 state of holes in the background of one filled Landau level. In this picture the 2/3 state is surrounded by a $\nu=1$ state at the boundary, which in turn is surrounded by vacuum [\onlinecite{MacDonaldTwoThird,YangTwoThird}]. The edge between 1 and vacuum supports a forward moving mode, whereas the edge between 2/3 and 1 supports a backward moving mode.  The physics suggested by the CF theory has similarity to the picture described above. At filling factor 2 we have two edges, one separating 2 and 1, and the other between 1 and 0 ({\em i.e.} vacuum). Upon antiparallel flux attachment, filling factors 2, 1 and 0 turn into 2/3, 1, and 0, thus again producing a $\nu=1$ region separating the 2/3 state and the vacuum. One can therefore expect a counter-propagating mode by the reasoning given above.

An analogous picture is not available for the spin singlet 2/3 state, however. This state cannot be viewed as the hole partner of any principal state, because particle hole symmetry in the presence of spin relates $\nu$ to $2-\nu$. The picture for the edge of the spin unpolarized 2/3 state is qualitatively different. The edge of the spin unpolarized state at 2 goes directly from 2 to 0, implying a 2/3-0 edge for the spin unpolarized 2/3 state. Because of the absence of $\nu=1$ at the boundary, it is not obvious why there should be an up-stream edge mode.

Nonetheless, an effective $K$ matrix description of Wen [\onlinecite{WenReview}] indicates a backward moving edge mode for both spin unpolarized and spin polarized states. The $K$ matrix can be obtained in the CF basis straightforwardly by noting that the 2$\times$2 $K$ matrix for filling factor 2 is $K_{jk}=-\delta_{jk}$ for antiparallel field. Composite fermionization of electrons by attachment of two vortices amounts to adding 2 to each each element of $K$, giving 
\begin{equation}
K_{2/3}=
\begin{pmatrix} 
1  & 2 \\ 2 & 1
\end{pmatrix}
\end{equation}
for both the spin unpolarized and spin polarized 2/3 states. This matrix has one positive and one negative eigenvalue, implying one down-stream and one up-stream edge mode. As explained in Ref. [\onlinecite{McDonald}], this structure of $K_{2/3}$ possesses a hidden $SU(2)$ symmetry, with an $SU(2)$ algebra generated by the neutral modes. For contrast, the $K$ matrix of the 2/5 state in the CF basis is given by 
\begin{equation}
K_{2/5}=
\begin{pmatrix} 
3  & 2 \\ 2 & 3
\end{pmatrix},
\end{equation}
obtained by adding 2 to each element of $\delta_{jk}$; both eigenvalues of this matrix are positive, hence no backward moving modes (neglecting edge reconstruction).

Yet another way to intuitively understand why there should be backward moving modes in the $n/(2n-1)$ FQH state is to note that composite fermions experience a negative effective field. Because switching the direction of the magnetic field reverses the direction of the $ {\bf E} \times {\bf B}$ drift, we can expect composite fermions at the edge moving in the backward direction.  More explicitly, consider the wave function in Eq.~(\ref{ReverseFlux}) below, written in the symmetric gauge. The edge modes at $n/(2n-1)$ derive from the edge modes of $\Phi_n$. The state $\Phi_n$ has $n$ independent edge modes, and the energy of a single excitation at each edge increases with its angular momentum (relative to the ground state). However, because of the complex conjugation of $\Phi_n$, an increase in the angular momentum in $\Phi_n$ translates into a {\em decrease} in the angular momentum at $n/(2n-1)$, thus producing a mode moving in the opposite direction. While this seems to give a rather nice picture for the origin of backward moving modes, it suggests that there are as many backward moving modes at $n/(2n-1)$ as there are forward moving modes at $n$, which is inconsistent with the effective $K$ matrix description that produces a single backward moving edge mode. 

Our aim in this work is to gain a microscopic understanding of the edge excitations of the negative flux CF states by considering the example of 2/3, and to bring consistency between the different approaches. We show, by an explicit construction of the wave functions, that only the neutral combination of the upstream edge modes survives at low energies, which is also in good agreement with exact diagonalization studies. In addition, we identify the forward moving modes with the so called ``Schur" modes, which are obtained by multiplying the ground state wave function by the symmetric Schur polynomials; these are analogous to the edge modes at filling factor one. We study both polarized and unpolarized states at 2/3, and, for comparison, also show results of the $\nu=2/5$ state, which is also described as $\nu=2$ filled $\Lambda$Ls but with parallel flux attachment.

The edge states of $\nu=2/3$ have also been studied by exact diagonalization. Johnson and MacDonald [\onlinecite{MacDonaldTwoThird}] and Hu {\em et al.} [\onlinecite{YangTwoThird}] model the spin polarized 2/3 state as 1/3 of holes inside a $\nu=1$ droplet. Moore and Haldane [\onlinecite{MooreTwoThird}] have studied the spin singlet 2/3 state by exact diagonalization to demonstrate the presence of a backward moving mode. The validity of the CF theory for electron droplets in the disk geometry, which can contain complex edges, has been studied extensively in a number of previous articles [\onlinecite{Kawamura}].

We note that the edge of the spin singlet 2/3 state provides a striking realization of spin charge separation, with pure spin and pure charge modes moving in opposite directions. This should in principle be observable. These modes have been labeled ``spinons" and ``chargeons" by Balatsky and Stone [\onlinecite{Balatsky}].

The paper is organized as follows. We introduce the CF wave functions and present the parton construction based on them in section II. Our model and numerical methods are briefly explained in section III. The energy spectra of spin-unpolarized $\nu=2$ and $2/5$ states are studied in section IV. The energy spectra of spin-unpolarized and polarized $\nu=2/3$ state are analyzed in section V and VI, respectively. We conclude in section VII. 

\section{Composite Fermion and Parton Construction}

In order to study the edge properties of a FQH state, we choose the disk geometry where the $n^{th}$ LL single particle states in the symmetric gauge are given by
\begin{equation} 
\eta_{n,m}(z)=\frac{(-1)^n}{\sqrt{2\pi}}\sqrt{\frac{n!}{2^m(m+n)!}} e^{-|z|^2/4} z^m L_n^m\left(\frac{|z|^2}{2}\right), 
\label{LLwave}
\end{equation}
where $L_n^m(x)$ is the associated Laguerre polynomial, $n$ and $m$ denote the LL index and angular momentum index respectively, $z=x-iy$ is the complex representation of electron coordinates, and all lengths are measured in units of the magnetic length $l$. The lowest Landau level (LLL) states ($n=0$) are of special importance and they are
\begin{equation}
\eta_{0,m}(z)=\frac{z^me^{-|z|^2/4}}{\sqrt{2\pi2^mm!}}
\label{LLLwave} 
\end{equation}
The wave function of the completely filled LLL is 
\begin{equation} 
\Phi_{1}(\{z_i\})=\prod_{i<j}(z_i-z_j).
\end{equation}
where we have omitted, for notational ease, the ubiquitous Gaussian factor and also the normalization coefficient.

In the CF theory, the system of strongly correlated electrons at filling factor given by Eq.~(\ref{fillfactor}) is mapped into a system of weakly interacting composite fermions at filling factor $\nu^*=n$. The wave function of this state is constructed as [\onlinecite{JainCF}]
\begin{equation}
\Psi_{n\over 2n+1}^{\rm gs} = {\cal P}_{\rm LLL} \left[\Phi_1 \Phi_1 \Phi_n\right]
\label{ParallelFlux}
\end{equation}
and 
\begin{equation}
\Psi_{n\over 2n-1}^{\rm gs} = {\cal P}_{\rm LLL} \left[\Phi_1 \Phi_1 \Phi_n^*\right]
\label{ReverseFlux}
\end{equation}
where $\Phi_n$ is a wave function at filling factor $\nu^*=n$. The first (second) of the above equation refers to situation when the effective magnetic field is parallel (antiparallel) to the external magnetic field. 
The spin degree of freedom is incorporated by assigning spins to the composite fermions [\onlinecite{Wu93,Park98}]. To form an incompressible state, the composite fermions of each spin species independently occupy an integer number $n_\uparrow$ and $n_\downarrow$ of $\Lambda$ levels, with $n=n_\uparrow+n_\downarrow$. $\Phi_n$ in the above equations is replaced by $\Phi_{n_\uparrow,n_\downarrow}$, where $n_\uparrow$ spin-up $\Lambda$Ls and $n_\downarrow$ spin-down $\Lambda$Ls are filled:
\begin{equation}
\Psi_{n\over 2n+1}^{\rm gs} = {\cal P}_{\rm LLL} \left[\Phi_1 \Phi_1\Phi_{n_\uparrow,n_\downarrow}\right], \;
\Psi_{n\over 2n-1}^{\rm gs} = {\cal P}_{\rm LLL} \left[\Phi_1 \Phi_1 \Phi_{n_\uparrow,n_\downarrow}^*\right]
\label{CFspin}
\end{equation}
In particular, for the spin-polarized $\nu=2/3$ and $2/5$ ground states considered below, the composite fermions are polarized and occupy the lowest two spin-up $\Lambda$Ls, whereas for the spin-unpolarized $\nu=2/3$ and $2/5$ ground states, both spin-up and spin-down composite fermions occupy one $\Lambda$L.

One can expect that the above wave functions also give a correspondence between the edge states at $\nu=n/(2n\pm 1)$ and $\nu^*=n$. In the disk geometry, the total angular momentum plays the role of the momentum. At $\nu^*=n$ there are $n$ edge modes, one corresponding to each LL, moving in the forward direction. They produce $n$ forward moving CF edge modes at $n/(2n+1)$, one for each $\Lambda$L, which is consistent with the description of the edge by other methods. However, for antiparallel flux attachment, they produce $n$ {\em backward} moving modes, because of the negative effective magnetic field for composite fermions, as indicated by the complex conjugation of $\Phi_n$ in Eq.~(\ref{ReverseFlux}). This simple view is in disagreement with the edge behavior from other methods. 

A more systematic description of the edge was developed by Wen [\onlinecite{WenParton}] using the parton model of the FQHE (Jain [\onlinecite{JainParton}]). In this model one imagines breaking each electron into an odd number of fictitious fermions, called partons, and describes an incompressible state as one in which each parton occupies an IQH state (which could be either parallel or antiparallel magnetic field). For the states of Eqs.~(\ref{ParallelFlux}) and (\ref{ReverseFlux}), we have three partons, individually occupying states with filling factor 1, 1, and $n$ (the last being in negative magnetic field for Eq.~(\ref{ReverseFlux})). Constraints on the charge and filling factors of various partons can be derived straightforwardly [\onlinecite{JainParton}]. If the partons are treated as independent, we have $n$ edge modes of the partons at filling $n$ (one from each LL), and one edge mode for each parton at filling one. However, the partons are obviously not independent degrees of freedom and must be identified in the calculation. Wen showed [\onlinecite{WenParton}] that imposing a constraint that annihilates all {\em relative} density oscillations produces an edge description that is consistent with the effective field theory description. This method can be straightforwardly applied to $n/(2n-1)$ states.

\begin{figure*}
\includegraphics[width=0.8\textwidth]{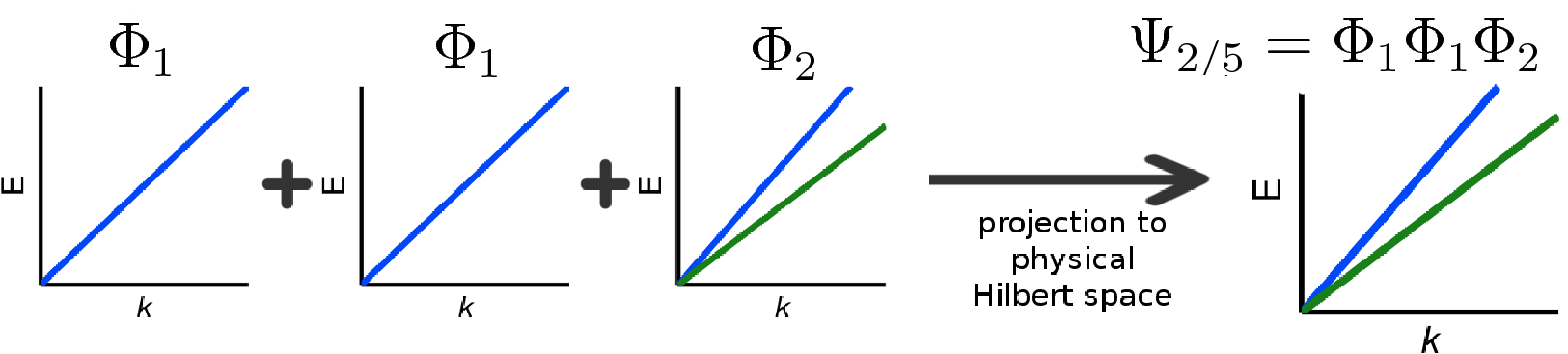}
\includegraphics[width=0.8\textwidth]{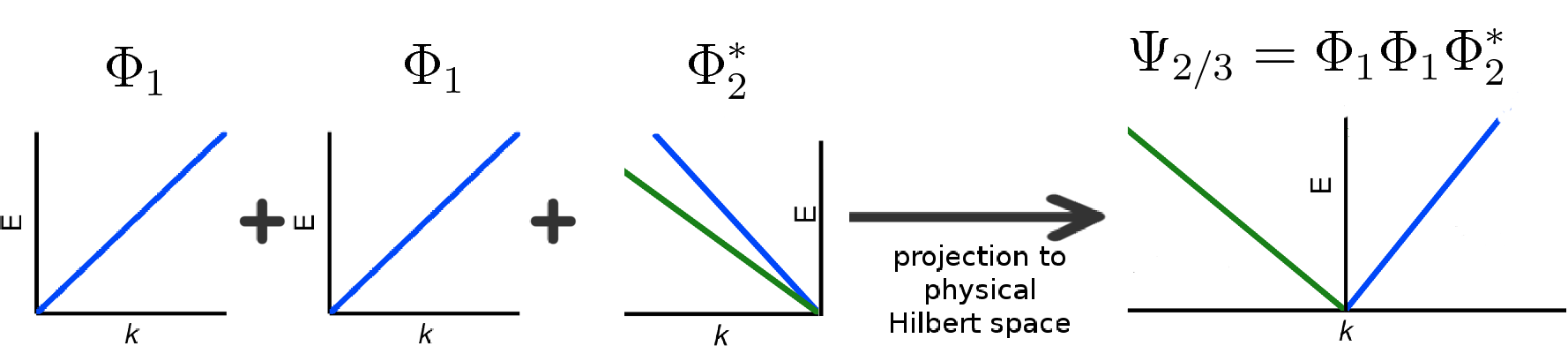}
\caption{(Color online) Schematics of the parton construction for the edge modes at $2/5$ and $2/3$. When spin is included, the symbol $\Phi_2$ is to be replaced by $\Phi_{2,0}$ for the spin polarized states and by $\Phi_{1,1}$ for the spin unpolarized states. The dispersions of various edge modes of the partons, and also of the edge modes of 2/5 and 2/3 after projection into the physical space.  The two colors represent charge (blue) and neutral (green) modes. $E$ and $k$ are the energy and wave vector.}
\label{PartonScheme}
\end{figure*}

We first briefly review the hydrodynamic approach for the edge physics [\onlinecite{WenReview}] of quantum Hall systems, which will be used below in the parton construction. Consider a Hall droplet with filling factor $\nu$. The electric field generated by the confining potential generates a current with speed $v=E/B$ along the edge, 
\begin{equation}
{\bf j}=\nu{\frac {e^2} h} \hat {\bf B} \times {\bf E}
\end{equation}
where $\hat {\bf B}$ is the direction of the magnetic field. The edge wave can be described by the one dimensional density $\rho(x)=nh(x)$ where $h(x)$ is the displacement of the edge, $x$ is the coordinate along the edge, and $n$ is the bulk electron density. The propagation of edge waves is described by 
\begin{equation}
\partial_t\rho \pm v\partial_x \rho=0
\label{waveprop}
\end{equation}
where the $+$ ($-$) sign applies when the magnetic field in the positive (negative) $z$ direction. 

The Hamiltonian of the edge wave is given by 
\begin{equation}
H=\int dx\ {\frac 1 2} eh\rho E=\int dx\ \pi {\frac v \nu} \rho^2
\end{equation}
In momentum space the wave equation and the Hamiltonian can be rewritten as
\begin{equation}
\dot \rho(k)= \pm i v k\rho(k)  \ \ \
H=2\pi {\frac v \nu} \sum_{k>0} \rho(k)\rho(-k)  \\
\end{equation}
where $\rho(k)=\int dx {\frac 1 {\sqrt{L}}} e^{ikx}\rho(x)$, and $L$ is the length
of the edge. Comparing with the standard Hamiltonian equation, $\rho_k|_{k>0}$ may be identified as the ``coordinates" and their corresponding canonical ``momenta" are $p(k)=\pm i 2\pi \rho(-k)/\nu k$. This theory is quantized by imposing the canonical commutation relation between $\rho(k)$ and $p(k)$, $[\rho(k),p(k^\prime)]=i\delta_{k k^\prime}$, which leads to the so-called $U(1)$ Kac-Moody algebra
\begin{equation}
[\rho(k), \rho(k^\prime)]={\pm \frac \nu {2\pi}}k\delta_{k+k'} \ \ \ \ k,k^\prime={\rm integer} \times {\frac {2\pi} L}
\label{KacMoody}
\end{equation}
In what follows, there will be several partons occupying one LL each; in that case, their density operators $\rho_\lambda$'s, with $\lambda$ labeling different partons, obey the commutation relation:
\begin{equation}
[\rho_\lambda(k), \rho_\mu(k')]={\pm \frac k {2\pi}}\delta_{\lambda \mu}\delta_{k+k'}.
\label{PartonCommute}
\end{equation}

\subsection{Spin-polarized states}

For fully spin polarized states, the ground state wave functions of $2/5$ and $2/3$ are given by Eq.~(\ref{ParallelFlux}) and (\ref{ReverseFlux}), where $\Phi_{n}$ is $\Phi_{2}$, the wave function of two filled spin-up $\Lambda$Ls. There are three types of partons which carry charges $2e/5$, $2e/5$ and $e/5$ for the 2/5 state, and $2e/3$, $2e/3$ and $-e/3$ for the 2/3 state [\onlinecite{JainParton}]. Following Wen [\onlinecite{WenParton}], we introduce density operators $\rho_1$, $\rho_2$, $\rho_3$ and $\rho_4$ where $\rho_{1,2}$ describe the edges of the two $\Phi_1$ state and $\rho_{3,4}$ describe the edges of the two filled LLs in $\Phi_2$ or $\Phi_2^*$. The commutators of $\rho_{1}$ and $\rho_{2}$ are given by Eq.~(\ref{PartonCommute}) with positive sign, and those of $\rho_{3}$ and $\rho_{4}$ with positive (negative) sign for the $2/5$ ($2/3$) state. To get a physical state of electrons from a state of partons, one must project away the unphysical degrees of freedom introduced through  the fictitious partons. For this purpose, we use the fact [\onlinecite{WenParton}] that the density fluctuations associated with ${\tilde \rho}_C=C_1\rho_1+C_2\rho_2+C_3(\rho_3+\rho_4)$ are unphysical for any $C_\alpha$ satisfying $\sum_{\alpha=1}^{3} C_\alpha=0$ and a physical operator must therefore commute with ${\tilde \rho}_C$  
\begin{equation}
[\hat O, {\tilde \rho}_C]=0
\label{PhysCommute}
\end{equation}
Before projecting to the physical Hilbert space, the edge excitations contain four branches described by the $\rho_\mu$'s. One can check that the following edge density operators commute with ${\tilde \rho}_C$
\begin{equation}
j_0=\sqrt{\frac 2 5}\left(\rho_0+{\frac 1 2}(\rho_3+\rho_4)\right) \ \ \ j_1=\sqrt{\frac 1 2}(\rho_3-\rho_4)
\end{equation}
for the $2/5$ state and
\begin{equation}
j_0=\sqrt{\frac 2 3}\left(\rho_0-{\frac 1 2}(\rho_3+\rho_4)\right) \ \ \ j_1=\sqrt{\frac 1 2}(\rho_3-\rho_4)
\end{equation}
for the $2/3$ state, where $\rho_0=\rho_1+\rho_2$. The physical edge excitations thus have two branches described by $j_0$ and $j_1$. This conclusion is consistent with the predictions of Chern-Simons theory [\onlinecite{WenReview}] and numerical calculations [\onlinecite{JainTwoFifth}]. We note that only $j_0$ couples to the external electric potential through $j_0 A_0$. The commutation relations between the $j_\mu$'s are
\begin{equation}
[j_\mu(k), j_\lambda(k')]={\pm \frac k {2\pi}} \delta_{\mu \lambda}\delta_{k+k'}
\label{FinalCommute}
\end{equation}
for $\mu, \lambda=0$ and $1$. The sign $\pm$ is $-$ only for the $j_1$ operator of the spin-polarized $2/3$ state, which describes a neutral backward moving edge mode. The parton construction is schematically shown in Fig.~\ref{PartonScheme} for 2/3 and 2/5.

\subsection{Spin-singlet states}

The CF ground state wave functions for spin-unpolarized $\nu=2/5$ and $2/3$ states are given by Eq.~(\ref{CFspin}), where the $\Phi_{n_\uparrow,n_\downarrow}$ is $\Phi_{1,1}$, the wave function of two filled LLs with spin-up and one spin-down. The parton construction for these states is analogous, again with four density operators defined as above. 
There are three types of partons for the unpolarized $\nu=2/5$ and $2/3$ states, which carry charges $2e/5$, $2e/5$ and $e/5$ and $2e/3$, $2e/3$ and $-e/3$ in these two cases. Four operators $\rho_1$, $\rho_2$, $\rho_3$ and $\rho_4$ are introduced to describe the edges of the parton states, where $\rho_{1,2}$ describe the edges of the two $\Phi_1$ state and $\rho_{c,s}$ describe the edges of $\Phi_{1,1}$ or $\Phi_{1,1}^{*}$, where $\rho_3$ and $\rho_4$ denote the density operators for spin-up and spin-down electrons of the $\nu=2$ unpolarized state. Following the same arguments as those used in the spin-polarized case, these operators satisfy the $U(1)$ Kac-Moody algebra Eq.~(\ref{PartonCommute}). It is convenient to combine these operators in a form that reflects the symmetry of the system under rotation in the spin space. 
Following Moore and Haldane [\onlinecite{MooreTwoThird}], we introduce operators $\rho_s=(\rho_3-\rho_4)/\sqrt{2}$ and $\rho_c=(\rho_3+\rho_4)/\sqrt{2}$ which commute with $S^2$ and $S_z$. They describe the spin and charge edge modes and their commutators are given by Eq.~(\ref{PartonCommute}). The density fluctuation operator is ${\tilde \rho}_C=C_1\rho_1+C_2\rho_2+\sqrt{2}C_3\rho_c$. The commutators of $\rho_{1,2}$ are given by Eq.~(\ref{PartonCommute}) with positive sign. The commutators of $\rho_{c,s}$ are given by Eq.~(\ref{PartonCommute}) with positive (negative) sign for the $2/5$ ($2/3$) state. A physical operator must commute with ${\tilde \rho}_C$ as shown in Eq.~(\ref{PhysCommute}). The following two sets of edge density operators are physical, 
\begin{equation}
j_c=\sqrt{\frac 2 5}\left(\rho_0+\frac {\rho_c} {\sqrt{2}}\right) \ \ \ j_s=\rho_s
\end{equation}
for the $2/5$ state and
\begin{equation}
j_c=\sqrt{\frac 2 3}\left(\rho_0-\frac {\rho_c} {\sqrt{2}}\right) \ \ \ j_s=\rho_s
\end{equation}
for the $2/3$ state where $\rho_0=\rho_1+\rho_2$. We note that only $j_c$ couples to the external electric potential through $j_c A_0$. The commutation relations between $j_\mu$'s are given by Eq.~(\ref{FinalCommute}) where the sign is $-$ only for the $j_s$ operator of the spin-unpolarized $2/3$ state, which describes a neutral spin mode moving in the backward direction. The parton constructions for spin singlet states are also schematically shown in Fig.~\ref{PartonScheme}.

\section{Model and Numerical Methods}

To test these ideas we have performed extensive numerical studies in various systems. In this section, we briefly explain our model and methods.

\subsection{Exact Diagonalization}

A semi-realisitc confinement potential [\onlinecite{YangEdgeRecon}] in the disk geometry can be modeled by a uniformly distributed positive charge background on a disk separated from the electron disk by a distance $d$. The Hamiltonian of such a system is 
\begin{eqnarray}
H &=& E_{\rm K}+V_{\rm ee}+V_{\rm eb}+V_{\rm bb}+E_Z \nonumber \\
&=& \sum_j \frac 1 {2m_b} \left(\vec{p}_j+\frac e c \vec{A}_j\right)^2 +
\sum_{j<k} \frac {e^2} {\epsilon |\vec{r}_j-\vec{r}_k|} \nonumber \\
&& - \rho_0\sum_{j} \int_{\Omega_N} d^2r \frac{e^2} {\epsilon \sqrt{|\vec{r}_j
-\vec{r}|^2+d^2}} \nonumber \\
&& + \rho_0^2 \int_{\Omega_N} \int_{\Omega_N} d^2r d^2r' \frac {e^2} {\epsilon 
|\vec{r}'-\vec{r}|} + g\mu_B B S_z
\end{eqnarray}
Here $m_b$ is the band mass of the electrons, $\vec{p}_j$ and $\vec{r}_j$ are the momentum and position operators of the $j$th electron, respectively. The quantity $\vec{A}_j$ is the vector potential of the magnetic field at $\vec{r}_j$, $\rho_0= \nu/2\pi l^2$ is the positive charge density spread over the background disk of radius $R_N$, and $\epsilon$ is the dielectric constant of the system. $S_z$ is the total spin in the $z$-direction. The $V_{\rm bb}$ term is a constant and does not affect the result, so we will drop it in what follows. It has been found that changing the distance $d$ can cause edge reconstruction [\onlinecite{YangEdgeRecon}], but the universal properties of edge states should not depend sensitively on the detailed nature of the confinement potential, so we also use a parabolic confinement potential give by $U(r)=\alpha r^2$ to simplify some calculations. Confining to the LLL and neglecting LL mixing, the Hamiltonian in the second quantized representation is given by
\begin{eqnarray}
H &=&\frac{1}{2}\sum_{r,s,t,u}\langle r,s|V_{\rm ee}|t,u\rangle a_r^\dagger 
a_s^\dagger a_t a_u \nonumber \\
&& +\sum_m \langle m|V_{\rm eb}|m\rangle a_m^\dagger a_m.
\end{eqnarray}
The two body electron-electron interaction coefficients and electron-background interaction coefficients are
\begin{equation}
\langle r,s|V_{\rm ee}|t,u\rangle=\int d^2r_1 d^2 r_2\eta^*_{r}(r_1)
\eta^*_{s}(r_2)\frac{e^2} {\epsilon r_{12}}\eta_{t}(r_1)\eta_{u}(r_2); 
\label{TwoBodyCoul}
\end{equation}
\begin{equation}
\langle m|V_{\rm eb}|m\rangle=-\rho_0 \int d^2r_1\int_{\Omega_N} d^2 r_2 
\frac{|\eta_m(r_1)|^2}{\sqrt{r_{12}^2+d^2}}
\label{EBInter}
\end{equation}
The structure of edge excitations of the $\nu=2/3$ spin-unpolarized state is not very apparent for the pure Coulomb interaction, and following Ref.~[\onlinecite{MooreTwoThird}] we consider a screened Coulomb interaction for which the edge excitations can be more readily identified. The two body matrix elements for screened Coulomb potential can be obtained by inserting a factor $\exp(-r^2/4 (\kappa l)^2)$ in Eq.~(\ref{TwoBodyCoul}), where $\kappa l$ is the screening length. 
The effect of parabolic confinement potential is very simple: it introduce an additive term $\beta M$ ($\beta=\hbar(\sqrt{(eB/m_b c)^2+8\alpha/m}-eB/m_b c)/2$) to the total energy of a state in the absence of confinement, where $M$ is the total angular momentum of the state. We shall shift the zero of the energy by changing this term to $\beta (M-M_0)$ where $M_0$ is the angular momentum of the ground state. The value of the confinement strength $\beta$ is chosen such that the state at $M_0$ becomes the ground state.

The Zeeman term $g \mu_B B S_z$ commutes with other terms in the Hamiltonian and thus can be considered separately. When this term is set to zero, the Hamiltonian commutes with the orbital and spin angular momentum operators, so the total orbital and spin angular momentum $M$ and $S^2$ are both good quantum numbers. We shall diagonalize in subspaces with fixed $M$ and $S_z=0$. The spin eigenvalue of a state can be calculated using 
\begin{equation}
{\hat S}^2={\hat S}_{-}{\hat S}_{+}+{{\hat S}_z}^2+{\hat S}_z
\end{equation}
The effect of the Zeeman term can be incorporated straightforwardly at the end of the calculation.

\subsection{Lowest Landau Level Projection}

In general, the wave function of composite fermions occupying $n$ $\Lambda$ levels, prior to LLL projection, is written as
\begin{equation}
\Psi(\{z\})=\text{Det} \left(
\begin{array}{ccc}
  \phi_1(z_1) & \cdots & \phi_1(z_N) \\
  \phi_2(z_1) & \cdots & \phi_2(z_N) \\
  \vdots & \cdots & \vdots \\
  \phi_N(z_1) & \cdots & \phi_N(z_N) \\
\end{array}
\right) \prod_{i<j}^N (z_i-z_j)^2
\end{equation}
where $\phi_i$'s are single particle states of the lowest $n$ LLs. This must be projected into the lowest LL to determine the low energy behavior.

There are two ways of performing the LLL projection. In the first one, first used by Dev and Jain  [\onlinecite{Wu93,DevJain}], the LLL projection is achieved by replacing the anti-holomophic coordinates $\bar z$ in the Slater determinant with $2 \partial/\partial z$. The evaluation of the projected wave function essentially amounts to expanding the unprojected wave function fully and then projecting each term into the LLL. The projection cannot be evaluated for a large number of electrons, because it requires keeping track of all basis states (Slater determinants) whose number grows exponentially with $N$. 

One can simplify the book-keeping by using Jack polynomials, which allows an efficient expansion of the Jastrow factor. We explain this briefly here. It has been shown that Jastrow factor belongs to a special class of polynomials, namely the Jacks [\onlinecite{FirstJack}]. The Schur function Eq.~(\ref{SchurPoly}) that will be used later is also a Jack. Since the single particle states in the LLL are indexed with angular momentum, a non-interacting $N$-particle state can be labeled by a partition $\lambda=\left[\lambda_1,\cdots,\lambda_N\right]$ in which the occupied single particle states are listed or a occupation configuration $n(\lambda)={n_m(\lambda),\ \ m=0,1,2,\cdots}$, where $m$ labels the single particle states and $n_m(\lambda)$ is the number of particles in the orbital $m$. The wave function corresponding to a partition is a monomial for bosons and a determinant for fermions. An interacting many body state is a superposition of many non-interacting basis states indexed by $\lambda$'s with coefficients $c_\lambda$. A squeezing operation for partitions is defined as follows: for a pair of particles in the orbitals $m_1$ and $m_2$, with $m_1<m_2-1$, the elementary squeezing operation consists of the two particles shifted to different momentum orbitals as $n_{m_{1,2}} \rightarrow n_{m_{1,2}}-1, n_{m_{1,2}\pm 1}  \rightarrow n_{m_{1,2}\pm 1}+1$. This means that both particles in the $m_1$, $m_2$ orbitals are shifted inwards. A partition $\lambda$ is said to dominate $\mu$ ($\lambda>\mu$) if $\mu$ can be generated by squeezing $\lambda$. A bosonic Jack can be expanded in terms of symmetric monomials 
\begin{equation}
J_{\lambda}^{\alpha}=\sum_{\kappa\le\lambda}c_{\lambda \kappa}(\alpha)\mathcal{M}_\kappa,
\end{equation}
where $\kappa$ runs over all partitions squeezed from the root partition $\lambda$ and $\mathcal{M}_\kappa$ is a monomial [\onlinecite{FirstJack}]. The root partition of the Jastrow factor of $N$ electrons, $\prod_{i<j}^{N}(z_i-z_j)^2$, is $\left[2N,2N-2,\cdots,0\right]$ and $\alpha=-2$. There is a recusive relation [\onlinecite{LongJack}] for the expansion coefficients $ c_{\lambda \kappa}(\alpha)$
\begin{equation}
c_{\lambda \kappa}(\alpha)=\frac{2/\alpha}{\rho_{\lambda}(\alpha)-\rho_{\kappa}(\alpha)} \sum_{\kappa<\mu\le\lambda}\hspace{-5pt}\left((l_i+t)-(l_j-t)\right)c_{\mu \kappa}(\alpha),
\label{JackExpan}
\end{equation}
The sum in Eq.~(\ref{JackExpan}) extends over all partitions $\mu$ strictly dominating $\kappa$ but being dominated or equal to $\lambda$. The $\rho$'s are defined as:
\begin{equation}
\rho_{\lambda}(\alpha)=\sum_i \lambda_i \left(\lambda_i -1 - \frac{2}{\alpha}(i-1)\right).
\end{equation}
Once the expansion is obtained, one can act the derivative on each monomial and sort the results to Slater determinants, \emph{i.e.}, Fock states of fermions. This method simplifies the calculation, but the computational time still grows exponentially and cannot be used for large systems; this projection is typically not possible beyond 10 particles. 

The second method is the Jain-Kamilla projection [\onlinecite{JainKamilla}]. In this method we absorb the Jastrow factor into the Slater determinant to write
\begin{equation}
\Psi(\{z\})=\text{Det} \left(
\begin{array}{ccc}
  \phi_1(z_1) J_1 & \cdots & \phi_1(z_N) J_N\\
  \phi_2(z_1) J_1 & \cdots & \phi_2(z_N) J_N\\
  \vdots & \cdots & \vdots \\
  \phi_N(z_1) J_1 & \cdots & \phi_N(z_N) J_N\\
\end{array}
\right)
\label{JKP}
\end{equation}
where $J_i=\prod_{k}^{\prime}(z_i-z_k)$ and the summation runs over all indices $k\neq i$. Instead of applying ${\cal P}_{\rm LLL}$ to the whole expression, one apply it to each matrix element individually and then evaluate the determinant. This method does not require decomposition of the wave function in the Slater determinant basis, and thus can be applied to very large systems for both parallel and antiparallel flux attachments [\onlinecite{JainKamilla},\onlinecite{SimonP}].

The two methods for projection do not produce identical wave functions. However, explicit calculations have shown that they are very close for fully spin polarized states. The spin unpolarized states are somewhat more sensitive to which projection is used, and we have found that the states obtained from the Dev-Jain projection are closer to the exact Coulomb states. In our calculations below, the Jain-Kamilla projection has been used for the spin-polarized states, and Dev-Jain projection for the spin-unpolarized states. 

\begin{table}[t]
\begin{center}
\begin{tabular}{c|c|c|c}
$\Delta M$ & $S_z=0$ & $S_z=\pm 1$ & $S_z=\pm 2$ \\
\hline
0 & 1 & 0 & 0\\
1 & 2 & 1 & 0\\
2 & 5 & 2 & 0\\
3 & 10 & 5 & 0\\
4 & 20 & 10 & 1\\
5 & 36 & 20 & 2\\
\end{tabular}
\end{center}
\caption{Number of all edge modes for various values of $\Delta M$ and $S_z$.}
\label{AllModeCount}
\end{table}

\begin{table}[t]
\begin{center}
\begin{tabular}{c|c|c|c}
$\Delta M$ & $S_z=0$ & $S_z=\pm 1$ & $S_z=\pm 2$ \\
\hline
0 & 1 & 0 & 0\\
1 & 1 & 1 & 0\\
2 & 2 & 1 & 0\\
3 & 3 & 2 & 0\\
4 & 5 & 3 & 1\\
5 & 7 & 5 & 2\\
\end{tabular}
\end{center}
\caption{Number of pure spin edge modes for various values of $\Delta M$ and $S_z$.}
\label{SpinModeCount}
\end{table}

\begin{table}[t]
\begin{center}
\begin{tabular}{c|c}
$\Delta M$ & $S_z=0$ \\
\hline
1 & 1 \\
2 & 2 \\
3 & 3 \\
4 & 5 \\
5 & 7 \\
\end{tabular}
\end{center}
\caption{Number of pure charge edge modes for several $\Delta M$. They all have $S_z=0$.}
\label{CharModeCount}
\end{table}

\begin{table}[t]
\begin{center}
\begin{tabular}{c|c}
$\Delta M$ & $S$\\
\hline
0 & 0\\
1 & 1\\
2 & 0,1\\
3 & 0,1,1\\
4 & 0,0,1,1,2\\
5 & 0,0,1,1,1,2,2\\
\end{tabular}
\end{center}
\caption{Number of pure spin edge modes for given $\Delta M$ and $S$.}
\label{SquareModeCount}
\end{table}

\section{Edge Modes of Spin-unpolarized $\nu=2$ and $\nu=2/5$ State}

Both the spin-unpolarized $2/3$ and 2/5 states are closely related to the spin-unpolarized $\nu=2$ state.  The edge modes at $\nu=2$ consist of one pure charge branch and one pure spin branch. Their counting can be obtained straightforwardly [\onlinecite{MooreTwoThird}]; the number of edge excitations in the subspaces with fixed $S_z$ values are shown in Table~\ref{AllModeCount} for some values of $\Delta M$. Note that a $S_z=\pm A$ state appears only if $\Delta M \geq A^2$.  Among these states, some are pure spin states, some are pure charge states, and some mixed. Tables~\ref{SpinModeCount} and~\ref{CharModeCount} show the number of pure spin and pure charge modes. Since the states form $SU(2)$ multiplets, the number of state in the $S^2=A(A+1)$ sector can be obtained by subtracting the number of state in the $S_z=A+1$ sector from that of the $S_z=A$ sector; some instances are summarized in Table~\ref{SquareModeCount}. We expect identical counting for 2/5.

\begin{figure}
\includegraphics[width=0.45\textwidth]{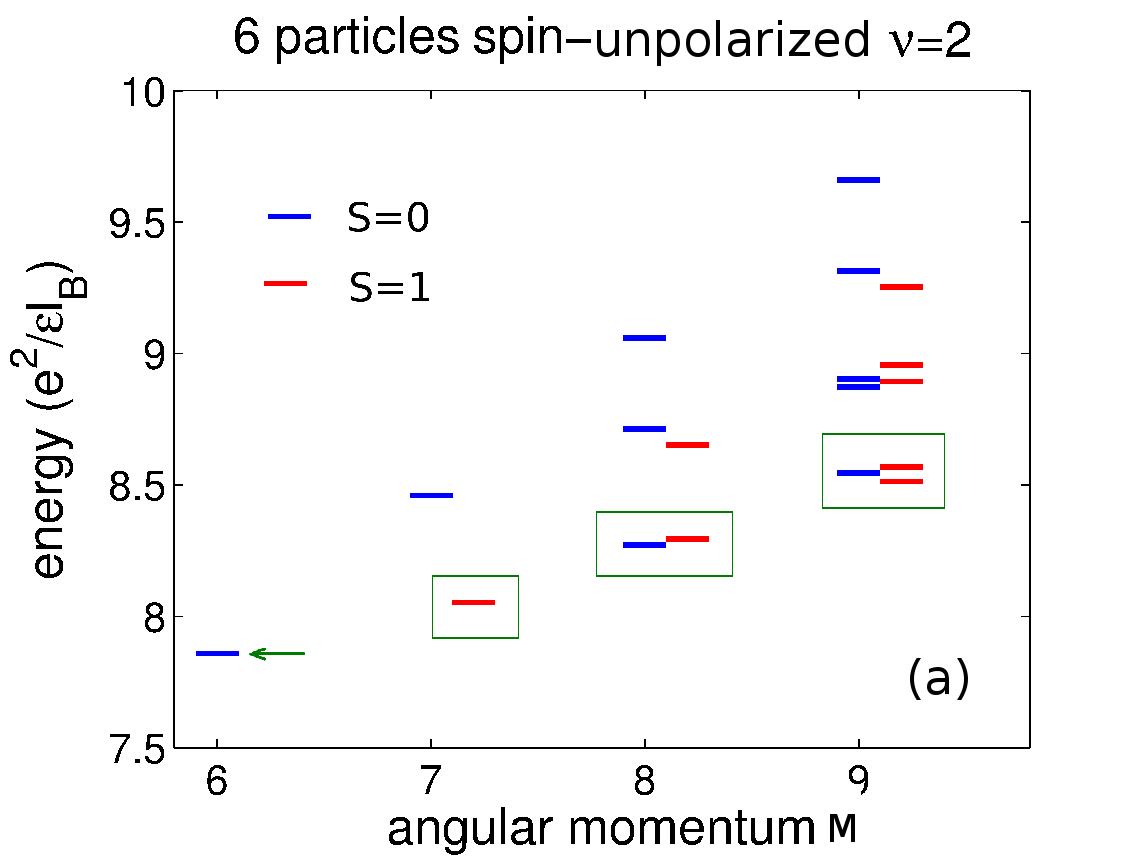}
\includegraphics[width=0.45\textwidth]{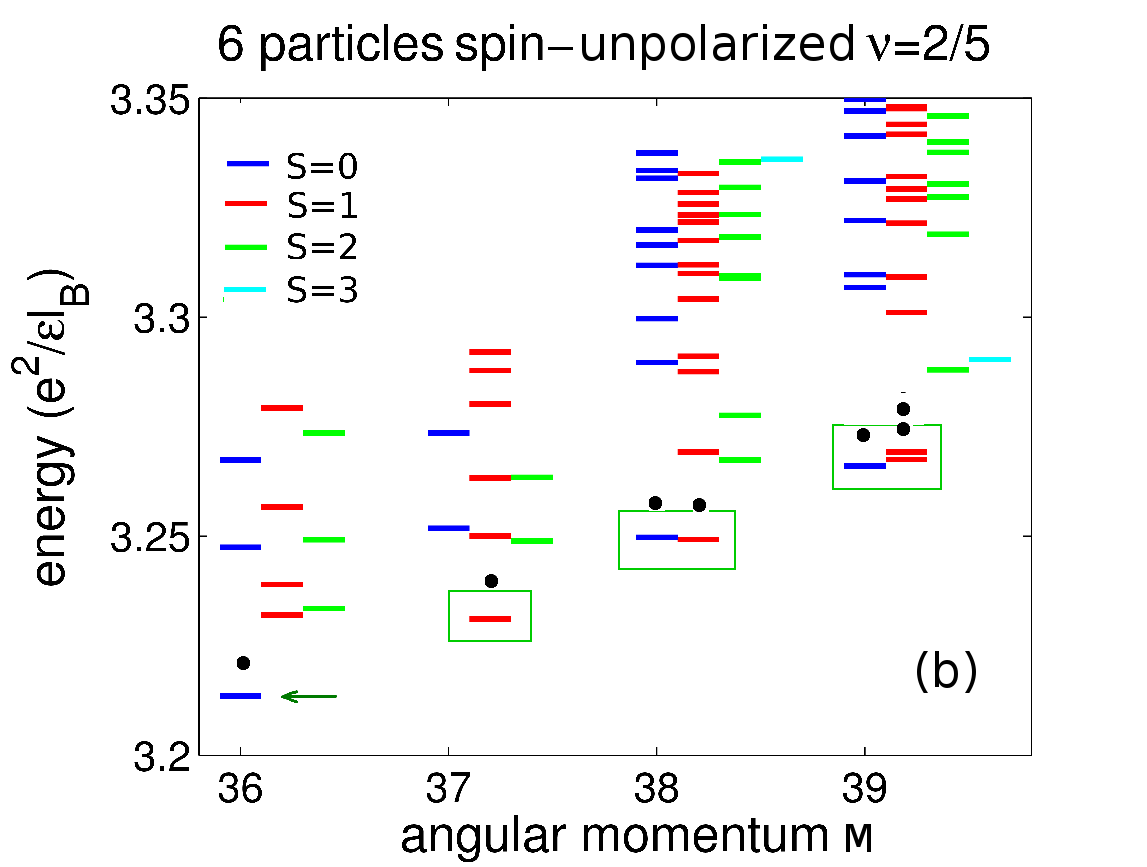}
\caption{(Color online) Pure spin excitations at $\nu=2$ and 2/5. Energy spectra of the edge excitations of the unpolarized $\nu=2$ (panel (a)) and $2/5$ (panel (b)) states with 6 particles. The ground states have $M=6$ and $M=36$, respectively. The pure spin modes are enclosed by green boxes, and the black dots in the lower panel show the energies of CF wave functions. The full spectrum at $\nu=2$ consists of pure spin, pure charge (Fig.~\ref{TwoCharge}), and mixed excitations. At $\nu=2/5$, the spectrum also includes excitations in the interior of the system where composite fermions are excited across $\Lambda$Ls. In this and the subsequent figures, eigenstates with different spin quantum number are shown in different colors, with the color coding indicated on the figures, and also horizontally shifted for clarity.}
\label{TwoSpin}
\end{figure}

\begin{figure}
\includegraphics[width=0.45\textwidth]{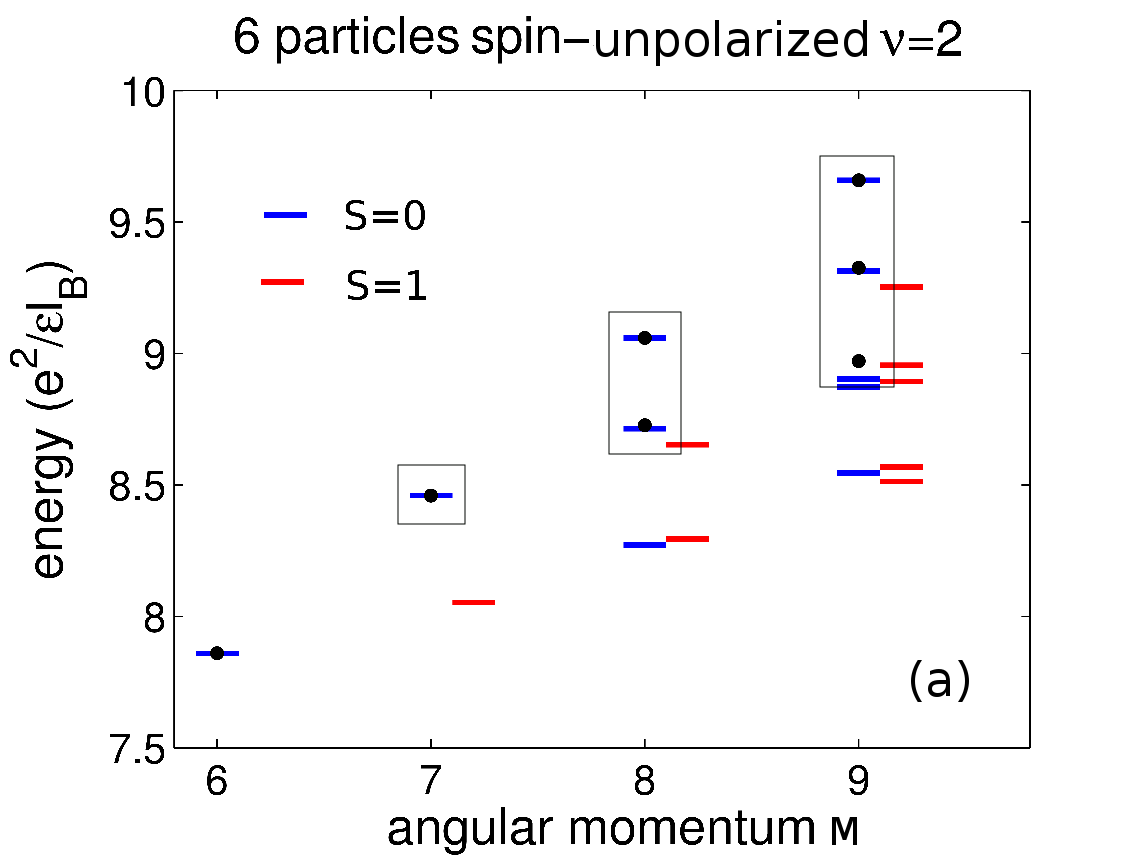}
\includegraphics[width=0.45\textwidth]{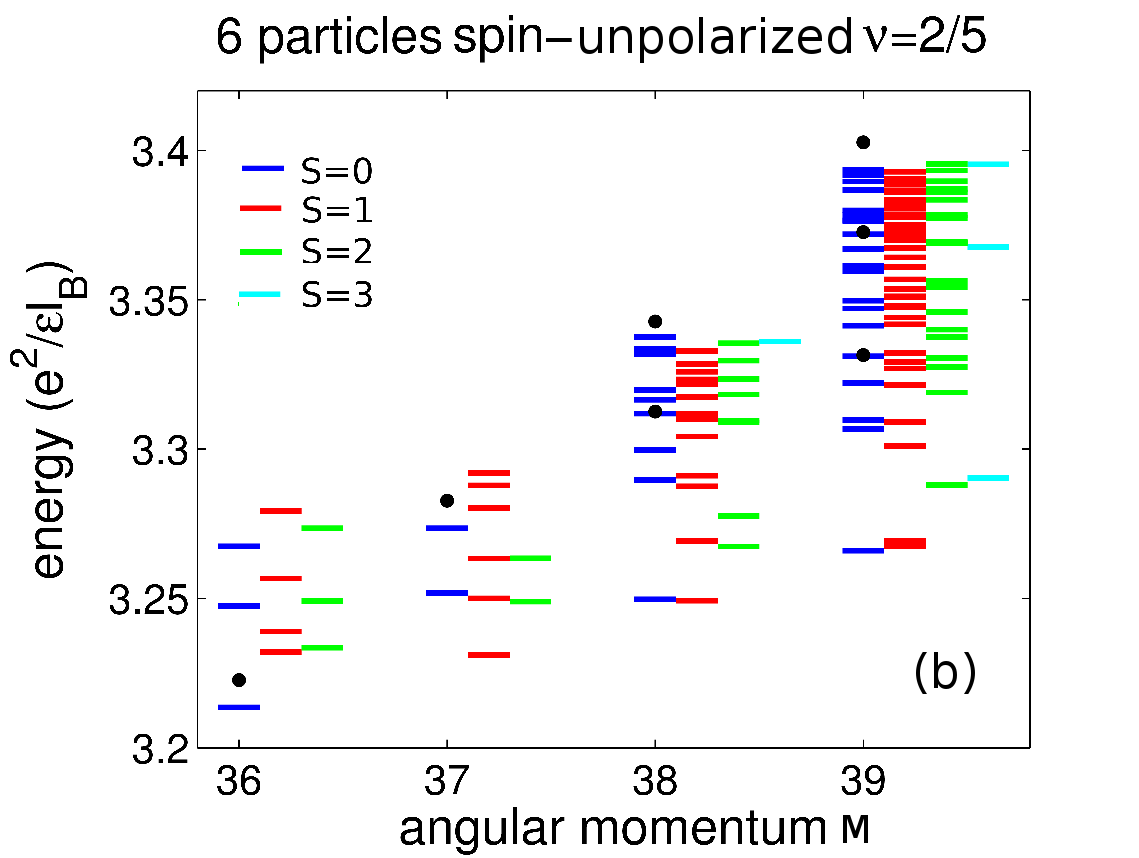}
\caption{(Color online) Pure charge excitations at $\nu=2$ and 2/5. Energy spectra of the unpolarized $\nu=2$ (panel (a)) and $2/5$ (panel (b)) states with 6 particles. The ground states have $M=6$ and $M=36$, respectively. The pure charge excitations of $\nu=2$ are enclosed by grey boxes. The pure charge excitations at $\nu=2/5$ are harder to identify for small systems because they lie in the continuum of the bulk excitations.}
\label{TwoCharge}
\end{figure}

Fig.~\ref{TwoSpin} shows the spectra for $\nu=2$ and $\nu=2/5$ states with the Zeeman energy set to zero. The coefficient $\beta$ due to the parabolic confinement potential is chosen to be $0.6$ and $0.06$ for the $\nu=2$ and $2/5$ state, respectively, so as to make the compact state the ground state. The energy eigenstates are also $S^2$ eigenstates., and we use different colors to represent different $S^2$ eigenvalues, and also shift the energy levels in the horizontal direction for clarity. The ground states are marked by green arrows. We see in both spectra low energy states (enclosed by green boxes), which we identify as pure spin edge states; these are well separated from other states. 
This counting matches that of the pure spin mode as shown in Table~\ref{SquareModeCount}. We introduce the following operators to describe the edge modes of the $\nu=2$ spin-unpolarized state (the superscripts ``c" and ``s" refer to ``pure charge" and ``pure spin" respectively):
\begin{equation}
C_m^{\dagger,\text {c}}=\sum_{n} a_{\uparrow m+n}^\dagger a_{\uparrow n} + a_{\downarrow m+n}^\dagger a_{\downarrow n}
\end{equation}
\begin{equation}
C_{m,0}^{\dagger,\text {s}}=\sum_{n} a_{\uparrow m+n}^\dagger a_{\uparrow n} - a_{\downarrow m+n}^\dagger a_{\downarrow n}
\end{equation}
\begin{equation}
C_{m,+1}^{\dagger,\text {s}}=\sum_{n} a_{\uparrow m+n}^\dagger a_{\downarrow n}
\end{equation}
\begin{equation}
C_{m,-1}^{\dagger,\text {s}}=\sum_{n} a_{\downarrow m+n}^\dagger a_{\uparrow n}
\end{equation}
where $a_{\sigma m}$ is the annihilation operator for an electron with spin $\sigma$ and angular momentum $m$. They have the following commutation relations with the spin operators $S^{-}$ and $S^{+}$:
\begin{equation}
\left[S^{+},C_{m,0}^{\dagger,\text {s}}\right]=C_{m,1}^{\dagger,\text {s}}
\end{equation}
\begin{equation}
\left[S^{-},C_{m,1}^{\dagger,\text {s}}\right]=C_{m,0}^{\dagger,\text {s}}
\end{equation}
\begin{equation}
\left[S^{-},C_{m,0}^{\dagger,\text {s}}\right]=C_{m,-1}^{\dagger,\text {s}}
\end{equation}
and all other commutators vanish. 
The states obtained by acting these operators on the ground state are in general not eigenstates of the Coulomb Hamiltonian. However, the two appear to be adiabatically connected. We can construct a model Hamiltonian $H^{\text {c}}=\sum_{m}C_m^{\dagger,\text {c}} C_m^{\text {c}}$ for which the pure spin excitations appear as zero modes. These zero mode states can be obtained by acting $C_{m,0}^{\dagger,\text {s}}$ operators on the ground state, and are adiabatically connected to the spin edge modes in the $\nu=2$ spectrum with Coulomb interaction. 

Due to the strongly interacting nature of the state at $\nu=2/5$, it is not possible to construct similar operators explicitly, but the trial 
wave functions for the the ground state and also excitations of the unpolarized $\nu=2/5$ state can be obtained by composite fermionizing the corresponding states at $\nu=2$. The edge excitations of unpolarized $2/5$ state can be obtained via
\begin{equation}
\Psi_{2/5}^{\Delta M} = {\cal P}_{\rm LLL} \left[\Phi_1^2 \Phi_{1,1}^{\Delta M}\right]
\label{TwoFifthUnpolarEdge}
\end{equation}
where $\Phi_{1,1}^{\Delta M}$ is an edge state of the unpolarized $\nu=2$ state with angular momentum $\Delta M$ relative to its ground state. The pure spin modes of the unpolarized $\nu=2/5$ state can be obtained from the $\nu=2$ pure spin modes in this way, and we show the energies of such CF state using black dots in panel (b) of Fig.~\ref{TwoSpin}. The lowest LL projection has been performed by the method in Ref.~[\onlinecite{JainKamilla}].

How about the pure charge modes moving in the forward direction? We construct these modes by multiplying the ground state wave functions with Schur functions, which are symmetric polynomials defined as
\begin{equation}
\mathcal S_{\lambda^B}(\{z\})=\frac {{\text {Det}}_{\lambda^F} (\{z\})}{\prod_{i<j}^N(z_i-z_j)}
\label{SchurPoly}
\end{equation}
where $\lambda^B=\left[\lambda_1,\lambda_2,\cdots,\lambda_N\right]$ is a bosonic partition and ${\text {Det}}_{\lambda^F}(\{z\})$ is a Slater determinant with fermionic index $\lambda^F=\left[\lambda_1+N-1,\lambda_2+N-2,\cdots,\lambda_N\right]$. Multiplication by this function increases the angular momentum by $\Delta M=\sum_{i}^N \lambda^B_i$. The number of independent Schur functions at $\Delta M$ is equal to the number of partitions of integer $\Delta M$, which is consistent with the pure charge mode counting in Table~\ref{CharModeCount}. In Fig.~\ref{TwoCharge}, we show the comparison of Schur modes with exact states for unpolarized $\nu=2$ and $2/5$ states. At $\nu=2$ the Schur modes and exact states match very well. While all states shown in the $\nu=2$ spectrum are edge excitations (as excitations to higher LLs are suppressed), the exact $\nu=2/5$ spectrum in Fig.~\ref{TwoCharge} also contains bulk excitations. The pure charge modes are not clearly separated from the bulk states (in contrast to the pure spin modes discussed above), indicating a larger velocity for the pure charge mode. This is a finite size effect, however, and we expect that for large $N$ a well defined edge branch will appear.

\begin{figure}
\includegraphics[width=0.45\textwidth]{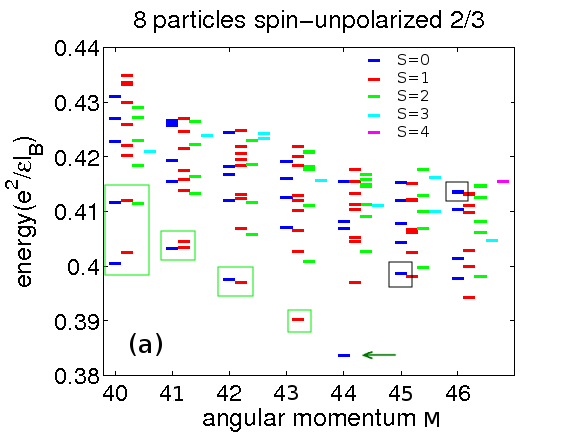}
\includegraphics[width=0.45\textwidth]{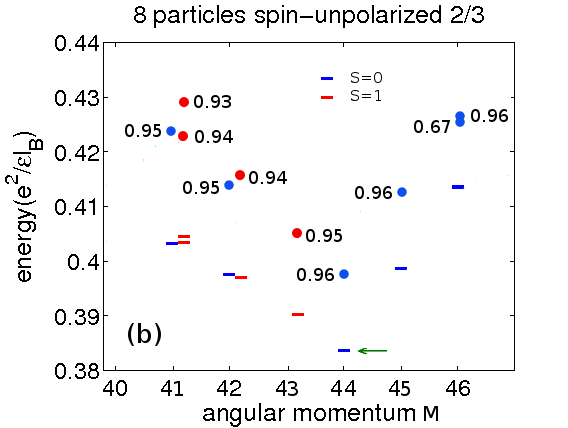}
\caption{(Color online) Energy spectrum of the edge excitations of the unpolarized $\nu=2/3$ state. The ground state, marked by a green arrow, occurs at total angular momentum $M=44$. In panel (a), backward moving pure spin modes are enclosed by green boxes and forward moving pure charge modes are enclosed by grey boxes. In panel (b), the dots show the energoes of the Schur states, and the nearby numbers show their overlaps with the exact states. For comparison, the pure spin and pure charge edge excitations from exact diagonalization spectrum of panel (a) are also shown in panel (b).}
\label{8PTwoThird}
\end{figure}

\begin{figure}
\includegraphics[width=0.45\textwidth]{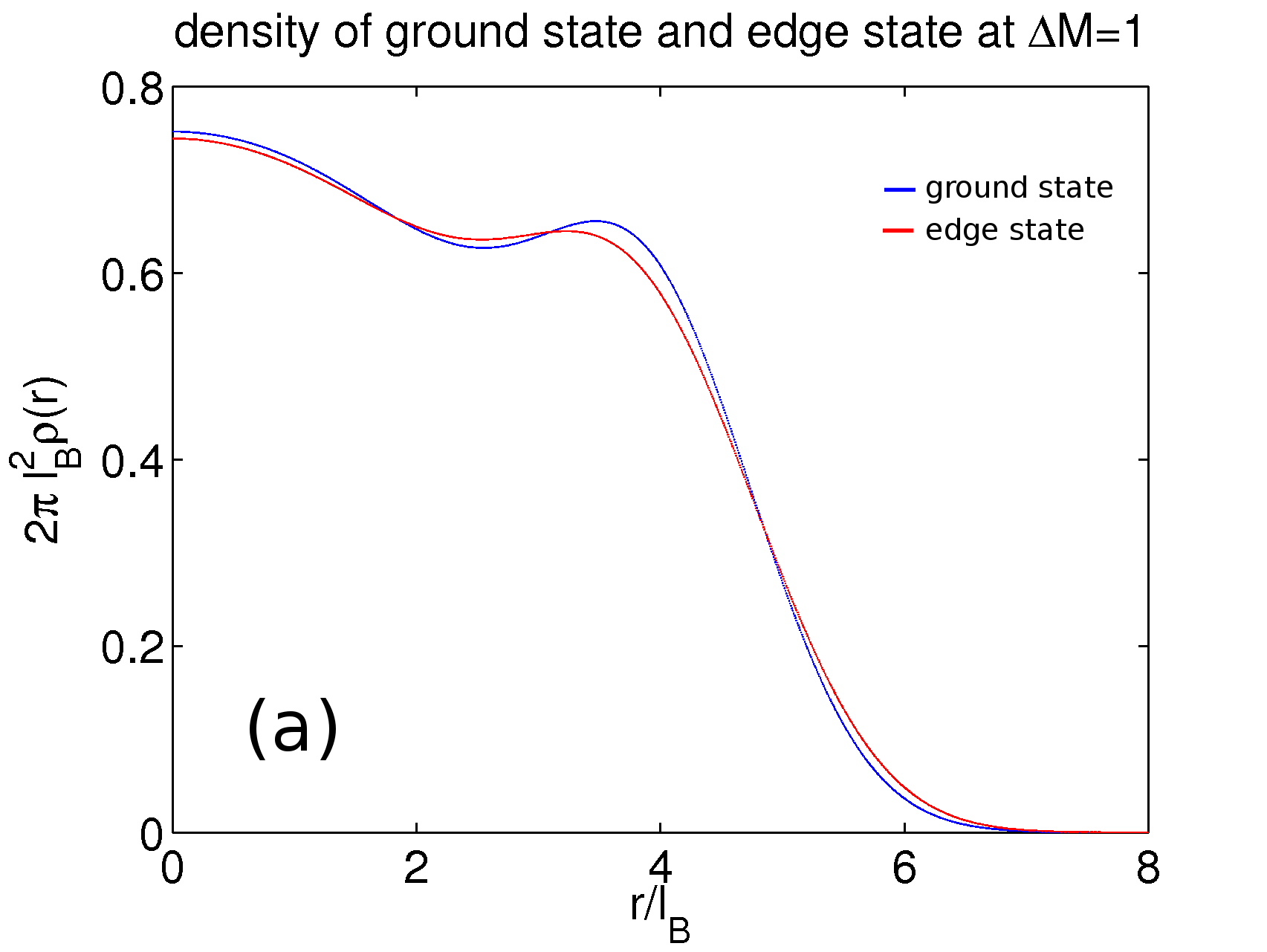}
\includegraphics[width=0.45\textwidth]{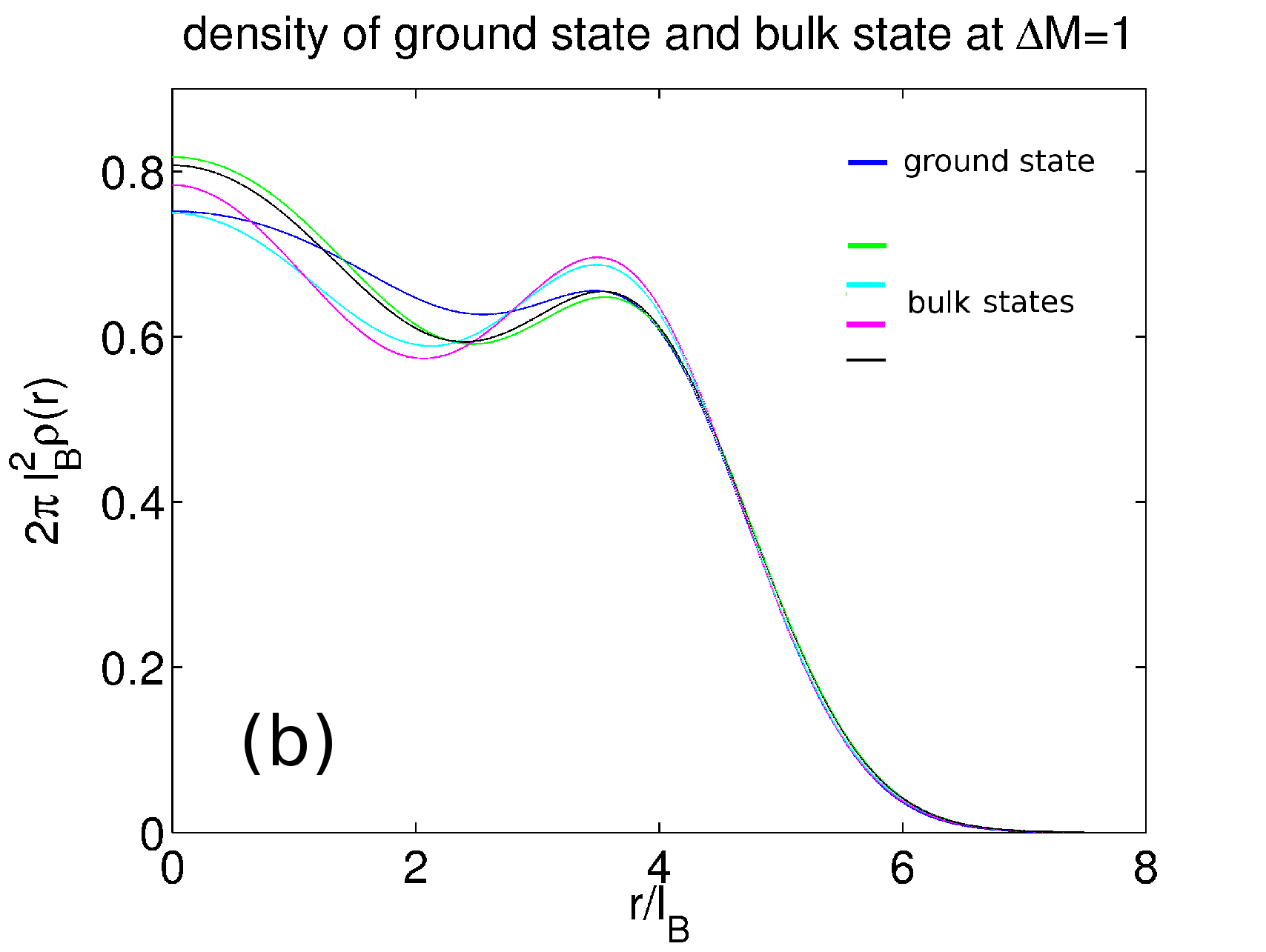}
\caption{(Color online) Density profiles of excited states identified as edge (panel (a)) and bulk (panel (b)) excitations of unpolarized $\nu=2/3$ state for $N=8$ at $\Delta M$=1 ($M=45$). For comparison, the density profile of the ground state at $M=44$ is also shown.}
\label{TwoThirdDensity}
\end{figure}

\section{Edge Modes of Spin-Unpolarized $\nu=2/3$ State}

We next come to the edge excitations of 2/3, where we expect counter propagating modes. In panel (a) of Fig.~\ref{8PTwoThird}, we show the spectra of spin-unpolarized $\nu=2/3$ with 8 particles for negative and positive $\Delta M$. The coefficient $\beta$ is 0.015 here. The ground state is marked by the green arrow. As explained before, we use a screened Coulomb interaction with screening length $\kappa l=l$ here. For negative $\Delta M$, some states, marked by the green boxes in panel (a) of Fig.~\ref{8PTwoThird}, are well separated from others. The counting of these states suggests that they are pure spin modes. It is not possible to construct explicit operators that would create the edge excitations of the unpolarized $\nu=2/3$ state, but trial wave functions can be obtained in CF theory using
\begin{equation}
\Psi_{2/3}^{-\Delta M} = {\cal P}_{\rm LLL} \left[\Phi_1^2 (\Phi_{1,1}^{\Delta M})^*\right]
\label{TwoThirdUnpolarEdge}
\end{equation}
We note {\em both} the forward moving edge modes of the unpolarized $\nu=2$ state are converted to backward moving modes of the $\nu=2/3$ state according to this transformation; because of the complex conjugation the factor $(\Phi_{1,1}^{\Delta M})^*$ contributes a negative angular momentum. Thus it appears that there would be two backward moving branches. However, we find that some CF states are annihilated by the LLL projection (evaluated by the method of Ref.~[\onlinecite{DevJain}], and among the surviving ones, many are pushed to higher energies.  The counting of the remaining low energy states matches with that predicted by the bosonized theory of the edge within the parton description. The black dots and nearby numbers in panel (b) of Fig.~\ref{8PTwoThird} show the energies of the CF trial states and their overlaps with the exact states enclosed by green boxes in panel (a). These results demonstrate that the CF theory captures the qualitative behavior, and while the agreement is quantitatively not as good as it is for the $n/(2n+1)$ states, it clearly gives a semiquantitative account of the backward moving edge.  The CF states with high energies (typically higher than the range of Fig.~\ref{8PTwoThird}) are not shown. 

For positive $\Delta M$, it is theoretically predicted that there are pure charge ({\em i.e.} $S=0$) excitations. In numerical calculations, we cannot identify any states well separated from the others at positive $\Delta M$'s and the states with lowest energies do not have $S=0$. This is to be expected, however, as the charge modes have higher velocity and therefore rapidly merge into the bulk excitations. 
By analogy to the discussion of the pure charge modes of the unpolarized $\nu=2$ and $2/5$ states, we expect that the pure charge modes of the $2/3$ state are also Schur modes. In panel (a) of Fig.~\ref{8PTwoThird}, the states enclosed by grey boxes are identified as pure charge modes. The red dots and nearby numbers in panel (b) show the energies of the Schur modes and their overlaps with these exact states. We have also calculated the energy spectrum of 6 particles. The energy differences between these Schur states and the lowest energy state decreases as the system size increases, which shows that the Schur modes will become the lowest energy states in the thermodynamic limit. To further support our identification of the edge and bulk excitations, we plot the density profiles of some states at $\Delta M=1$ in Fig.~\ref{TwoThirdDensity}; the state shown in panel (a) only exhibits density variations in the vicinity of the edge, whereas those in panel (b) deviate also in the bulk. Similar behavior is confirmed at $\Delta M=2$.

\begin{figure}
\includegraphics[width=0.45\textwidth]{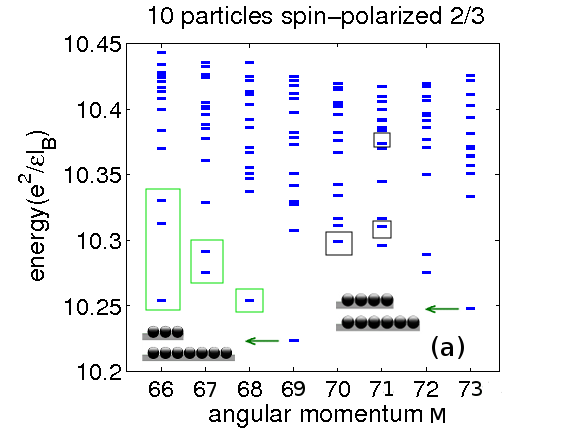}
\includegraphics[width=0.45\textwidth]{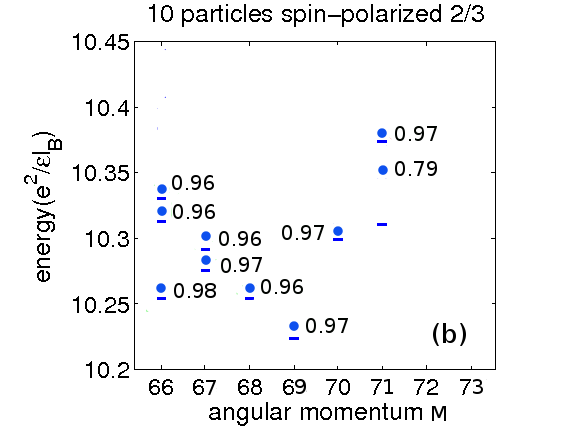}
\caption{(Color online) Energy spectrum of the edge excitations of the spin polarized $\nu=2/3$ state. The ground state occurs at total angular momentum $M=69$. In panel (a), the backward moving neutral modes are enclosed by green boxes and the forward moving charge modes by grey boxes. The two insets show the CF configurations at $M=69$ and 73. In panel (b), the dots and nearby numbers show the energies of CF and Schur states and their overlap with exact states (\emph{i.e.} boxed states in panel (a)). The relatively poor comparison for $M=71$ is attributed to the fact that the low energy modes here can also be viewed as the backward moving modes emanating from the ground state at $M=73$, and the two description compete; this will not be an issue for larger systems.}
\label{10PTwoThird}
\end{figure}

\section{Edge Modes of Spin-Polarized $\nu=2/3$ State}

CF wave functions of the edge excitations of the spin polarized $\nu=2/3$ state are constructed from the edge excitations of the spin-polarized $\nu=2$ state via the mapping
\begin{equation}
\Psi_{2/3}^{-\Delta M} = {\cal P}_{\rm LLL} \left[\Phi_1^2 (\Phi_2^{\Delta M})^*\right]
\label{TwoThirdPolarEdge}
\end{equation}
where $\left(-\right)\Delta M$ is the angular momentum measured relative to their respective ground states. More details about the construction of such wave functions can be found in Ref.~[\onlinecite{JeonDisk}]. We denote the state with $N_1$ composite fermions in the lowest and $N_2$ composite fermions in the second $\Lambda$L by [$N_1$, $N_2$].  Note that in order to have two independent CF edge branches, we should choose the number of composite fermions in the second $\Lambda$Ls to be sufficiently smaller than that in the lowest $\Lambda$ level, so as to eliminate transitions of composite fermions from the $2^{nd}$ $\Lambda$L to the $1^{st}$ $\Lambda$L (which would happen only at a large $\Delta M$). This also corresponds to the experimental situation where for a typical confinement potential we expect that the lowest $\Lambda$ level would extend farther than the second. (Note that the spin polarized 2/3 is different from the spin unpolarized 2/3 in this respect.) We show the energy spectrum of spin-polarized $2/3$ state with 10 particles in panel (a) of Fig.~\ref{10PTwoThird}. The coefficient $\beta$ is chosen to be 0.075 here. The ground state $[7,3]$ occurs at angular momentum $M=69$; it is shown by the left inset of Fig.~\ref{10PTwoThird}. The backward moving edge states are enclosed by green boxes. By construction, all the forward-moving $\nu^*=2$ edge states are transformed to backward-moving $\nu=2/3$ edge states, and one may expect two backward moving modes. We find that, similarly to the spin-unpolarized $\nu=2/3$ state, some of the CF states are projected out and some are pushed to high energies (typically outside the range of Fig.~\ref{10PTwoThird}, leaving only a few at low energies, which match very well with exact states. A comparison with the exact states is shown  in panel (b) of Fig.~\ref{10PTwoThird}, where black dots show the energies of the CF states and the numbers are the overlaps.

It is again natural to associate the forward moving mode with positive $\Delta M$ with Schur excitations. We again encounter the finite size difficulty of the absence of a clear gap in some cases. The red dots and nearby numbers in panel (b) of Fig.~\ref{10PTwoThird} show the energies of Schur modes and their overlaps with exact states. As we increase the angular momentum from 69 to 73, we get another CF configuration $[6,4]$, which can also serve as a finite size representation of the 2/3 state. That sets a finite size limitation on the angular momentum $\Delta M$ one can study in the forward or backward direction. For example, the state at angular momenta smaller than 73 may also be viewed as backward moving edge states emanating from $M=73$. This complicates the counting of the edge modes for finite size systems, and is also the likely cause of the mismatch between the exact spectra and the expectation from the effective theory.

\section{Concluding remarks}

We have studied the edge states of spin unpolarized and polarized $\nu=2/3$ states within the framework of the CF theory, both using parton construction and the microscopic CF wave functions. The parton construction of composite fermions produces one forward moving charge mode and one backward moving neutral mode for both spin unpolarized and polarized $2/3$ states, which agrees with the predictions of Chern-Simons effective field theory. Backward moving modes also appear naturally from the observation that at 2/3 composite fermions experience negative magnetic field, and thus move at the edge in a direction opposite to that of electrons. A naive counting would suggest {\em two} backward moving edge modes, one from each $\Lambda$ level, but we have shown, by explicit construction of the CF wave functions, that one mode is projected out of the low energy sector and the remaining excitations are good approximations of the exact states for both spin unpolarized and spin polarized 2/3 state. The forward moving modes are Schur modes; they are harder to identify in the exact spectra of small systems because, due to the larger velocity of these modes, they quickly enter into the continuum of bulk excitations across $\Lambda$ levels. Nonetheless, a careful examination of the density profiles has allowed us to identify the forward moving edge states and to compare them with Schur modes. We have thus shown that the description from the microscopic approach is consistent with the Chern-Simons or the parton approach, albeit only after a nontrivial reduction of the edge excitations upon projection into the low energy space.  Annihilation of mean field CF states upon projection has been found in previous numerical studies in other contexts as well [\onlinecite{DevJain,WuJain}], but no understanding exists of the general mathematical structure underlying such annihilations.

Before closing, we note that a number of effects have been left out in our study. While we have only focused on state counting in this paper, the effective description in terms of bosons also makes predictions for the spectral function, {\em i.e.}, matrix elements relating ground to excited states through the electron creation operator [\onlinecite{PalaciosMacDonald}], which we have not investigated. We have also not considered subtle questions regarding the antisymmetry of the electron operator in the projected edge state space [\onlinecite{Zulicke}], or the role of $\Lambda$L mixing. Similarly, the possibility of edge reconstruction [\onlinecite{YangEdgeRecon,PalaciosMacDonald,recon}] has not been incorporated into our calculations, which, if it occurs, will fundamentally alter the nature of the edge. The effects of finite thickness, LL mixing and disorder have also been neglected. 

\hspace{2em}

\emph{Acknowledgements}. We thank Arkadiusz W\'ojs and Shivakumar Jolad for help with numerical methods. Financial support from the DOE under grant no. SC0005042 is gratefully acknowledged.


\begin{thebibliography}{99}

\end{thebibliography}


\begin{thebibliography}{99}

\bibitem{IQHE} K. von Klitzing, G. Dorda and M. Pepper, Phys. Rev. Lett. {\bf 45}, 494 (1980).

\bibitem{FQHE} D. C. Tsui, H. L. Stormer, and A. C. Gossard, Phys. Rev. Lett. {\bf 48}, 1559 (1982).

\bibitem{JainCF} J. K. Jain, Phys. Rev. Lett. {\bf 63}, 199 (1989); Phys. Rev. B {\bf 41}, 7653 (1990).

\bibitem{Wu93} X. G. Wu, G. Dev and J. K. Jain, Phys. Rev. Lett. {\bf 71}, 153 (1993).

\bibitem{Park98} K. Park and J. K. Jain, Phys. Rev. Lett. {\bf 80}, 4237 (1998); Solid State Commun. {\bf 119}, 291 (2001).

\bibitem{SpinDegreeExp} 
R. R. Du \emph{et al.}, Phys. Rev. Lett. {\bf 75}, 3926 (1995); I. V. Kukushkin, K. von Klitzing, and K. Eberl, Phys.Rev. Lett. {\bf 82}, 3665 (1999).

\bibitem{FQHEdge} X. G. Wen, Phys. Rev. B {\bf 41}, 12838 (1990).

\bibitem{WenReview} X. G. Wen, Int. J. Mod. Phys. B {\bf 6}, 1711 (1992); Adv. in Phys. {\bf 44}, 405 (1995).

\bibitem{ChangReview} A. M. Chang, Rev. Mod. Phys. {\bf 75}, 1449 (2003).

\bibitem{Levitov} A. V. Shytov, L. S. Levitov, and B. I. Halperin, Phys. Rev. Lett. {\bf 80}, 141 (1998); L. S. Levitov, A. V. Shytov, and B. I. Halperin, Phys. Rev. B {\bf 64}, 075322 (2001).

\bibitem{FradkinSpin} A. Lopez and E. Fradkin, Phys. Rev. B {\bf 63}, 085306 (2001).

\bibitem{Kane} C.L. Kane, M. P. A. Fisher and J. Polchinski, Phys. Rev. Lett. {\bf 72}, 4129 (1994).

\bibitem{TwoThirdNeutral} A. Bid, N. Ofek, H. Inoue, M. Heiblum, C. L. Kane,	V. Umansky, and D. Mahalu, Nature {\bf 466}, 585 (2010).

\bibitem{MacDonaldTwoThird} M. D. Johnson and A. H. MacDonald, Phys. Rev. Lett. {\bf 67}, 2060 (1991).

\bibitem{YangTwoThird} Z.-X. Hu, H. Chen, K. Yang, E. H. Rezayi, and X. Wan, Phys. Rev. B {\bf 78}, 235315 (2008).

\bibitem{McDonald} I.A. McDonald and F.D.M. Haldane, Phys. Rev. B {\bf 53}, 15845 (1996).

\bibitem{MooreTwoThird} J. E. Moore and F. D. M. Haldane, Phys. Rev. B {\bf 55}, 7818 (1997).

\bibitem{Kawamura} J. K. Jain and T. Kawamura, Europhys. Lett. {\bf 29}, 321 (1995); G. S. Jeon, C.-C. Chang, and J. K. Jain, Eur. Phys. J. B {\bf 55}, 271 (2007); C. Shi, G. S. Jeon and J. K. Jain, Phys. Rev. B {\bf 75}, 165302 (2007).

\bibitem{Balatsky} A. Balatsky and M. Stone, Phys. Rev. B {\bf 43}, 8038 (1991).

\bibitem{WenParton} X. G. Wen, Mod. Phys. Lett. B {\bf 5}, 39 (1991).

\bibitem{JainParton} J. K. Jain, Phys. Rev. B {\bf 40}, 8079 (1989).

\bibitem{JainTwoFifth} G. J. Sreejith, S. Jolad, D. Sen, and J. K. Jain, Phys. Rev. B {\bf 84}, 245104 (2011).

\bibitem{YangEdgeRecon} X. Wan, Kun Yang, and E. H. Rezayi, Phys. Rev. Lett. {\bf 97}, 256804 (2006).

\bibitem{DevJain} G. Dev and J. K. Jain, Phys. Rev. B {\bf 45}, 1223  (1992);  Phys. Rev. Lett. {\bf 69}, 2843 (1992).

\bibitem{FirstJack} B. A. Bernevig and F. D. M. Haldane, Phys. Rev. Lett. {\bf 100}, 246802 (2008).

\bibitem{LongJack} R. Thomale, B. Estienne, N. Regnault, and B. A. Bernevig, Phys. Rev. B {\bf 84}, 045127 (2011).

\bibitem{JainKamilla} J. K. Jain and R. K. Kamilla, Phys. Rev. B {\bf 55}, R4895 (1997).

\bibitem{SimonP} G. M\"oller and S. H. Simon, Phys. Rev. B {\bf 72}, 045344 (2005); 
S. C. Davenport and S. H. Simon, Phys. Rev. B {\bf 85}, 245303 (2012). 

\bibitem{JeonDisk} G.-S. Jeon, C.-C. Chang and J. K. Jain, Eur. J. Phys. B {\bf 55}, 271 (2007).

\bibitem{WuJain} X.-G. Wu and J. K. Jain, Phys. Rev. B {\bf 51}, 1752 (1995).

\bibitem{PalaciosMacDonald} J. J. Palacios and A. H. MacDonald, Phys. Rev. 
Lett. {\bf 76}, 118 (1996); U. Z\"ulicke and A. H. MacDonald, Phys. Rev. B {\bf 54},
R8349 (1996); X. Wan, E. H. Rezayi, and K. Yang, Phys. Rev. B {\bf 68}, 
125307 (2003); X. Wan, K. Yang, and E. H. Rezayi, Phys. Rev. Lett. {\bf 88},
056802 (2002); S. Jolad and J. K. Jain, Phys. Rev. Lett {\bf 102}, 116801
(2009); S. Jolad, D. Sen and J. K. Jain, Phys. Rev. B {\bf 82}, 075315 (2010).

\bibitem{Zulicke} U. Z\"ulicke, J. J. Palacios, and A. H. MacDonald
Phys. Rev. B {\bf 67}, 045303 (2003).

\bibitem{MandalEdge} S. S. Mandal and J. K. Jain, Solid State Commun. {\bf 118}, 503 (2001);  Phys. Rev. Lett. {\bf 89}, 096801 (2002).

\bibitem{recon} C. de C. Chamon and X. G. Wen, Phys. Rev. B {\bf 49}, 8227
(1994); Y. N. Joglekar, H. K. Nguyen, and G. Murthy, Phys. Rev. B {\bf 68}, 035332 (2003).

\end{thebibliography}
\end{document}